\documentclass[manuscript, modern]{aastex63}
\received{September 10, 2019}
\revised{--}
\accepted{--}
\submitjournal{ApJ}

\shorttitle{Near-$f_{ce}$ Waves in the near-Sun Solar Wind}
\shortauthors{Malaspina et al.}
\begin{document}

\title{Plasma Waves near the Electron Cyclotron Frequency in the near-Sun Solar Wind}

\correspondingauthor{David M. Malaspina}
\email{David.Malaspina@colorado.edu}

\author[0000-0003-1191-1558]{David M. Malaspina}
\affiliation{Astrophysical and Planetary Sciences Department, University of Colorado, Boulder, CO, USA}
\affiliation{Laboratory for Atmospheric and Space Physics, University of Colorado, Boulder, CO, USA}

%%%% Co-authors (go alphabetical by instrument) : 
%%%% FIELDS Maker list
%%%% SWEAP Maker list
%%%% who worked on analysis (Bob)
%%%% who I discussed with (Vasko, Howes) - in Acknowledgements? 

\author[0000-0001-5258-6128]{Jasper Halekas}
\affiliation{University of Iowa, Iowa City, IA, USA}

\author[0000-0002-6075-1813]{Laura Ber\v{c}i\v{c}}
\affiliation{LESIA, Observatoire de Paris, PSL Research University, CNRS, Meudon, France}
\affiliation{Physics and Astronomy Department, University of Florence, Italy}

\author{Davin Larson}
\affiliation{Space Sciences Laboratory, University of California, Berkeley, CA, USA}

\author[0000-0002-7287-5098]{Phyllis Whittlesey}
\affiliation{Space Sciences Laboratory, University of California, Berkeley, CA, USA}

\author[0000-0002-1989-3596]{Stuart D. Bale}
\affiliation{Space Sciences Laboratory, University of California, Berkeley, CA, USA}
\affiliation{Physics Department, University of California, Berkeley, CA, USA}

\author[0000-0002-0675-7907]{John W. Bonnell}
\affiliation{Space Sciences Laboratory, University of California, Berkeley, CA, USA}

\author[0000-0002-4401-0943]{Thierry Dudok de Wit}
\affiliation{LPC2E, CNRS, and University of Orl\'eans, Orl\'eans, France}

\author{Robert E. Ergun}
\affiliation{Astrophysical and Planetary Sciences Department, University of Colorado, Boulder, CO, USA}
\affiliation{Laboratory for Atmospheric and Space Physics, University of Colorado, Boulder, CO, USA}

\author[0000-0003-1749-2665]{Gregory Howes}
\affiliation{University of Iowa, Iowa City, IA, USA}

\author[0000-0003-0420-3633]{Keith Goetz}
\affiliation{School of Physics and Astronomy, University of Minnesota, Minneapolis, MN, USA}

\author{Katherine Goodrich}
\affiliation{Space Sciences Laboratory, University of California, Berkeley, CA, USA}

\author[0000-0002-6938-0166]{Peter R. Harvey}
\affiliation{Space Sciences Laboratory, University of California, Berkeley, CA, USA}

\author[0000-0003-3112-4201]{Robert J. MacDowall}
\affiliation{NASA Goddard Space Flight Center, Greenbelt, MD, USA}

\author[0000-0002-1573-7457]{Marc Pulupa}
\affiliation{Space Sciences Laboratory, University of California, Berkeley, CA, USA}

%%%% And Now SWEAP 

\author{Anthony W. Case}
\affiliation{Harvard-Smithsonian Center for Astrophysics, Cambridge, MA, USA}

\author{Justin C. Kasper}
\affiliation{University of Michigan, Ann Arbor, MI, USA}

\author{Kelly E. Korreck}
\affiliation{Harvard-Smithsonian Center for Astrophysics, Cambridge, MA, USA}

\author{Roberto Livi}
\affiliation{Space Sciences Laboratory, University of California, Berkeley, CA, USA}

\author[0000-0002-7728-0085]{Michael L. Stevens}
\affiliation{Harvard-Smithsonian Center for Astrophysics, Cambridge, MA, USA}

\begin{abstract}

Data from the first two orbits of the Sun by Parker Solar Probe reveal that the solar wind sunward of 50 solar radii is replete with plasma waves and instabilities.  One of the most prominent plasma wave power enhancements in this region appears near the electron cyclotron frequency ($f_{ce}$).  Most of this wave power is concentrated in electric field fluctuations near 0.7 $f_{ce}$ and $f_{ce}$, with strong harmonics of both frequencies extending above $f_{ce}$.  At least two distinct, often concurrent, wave modes are observed, preliminarily identified as electrostatic whistler-mode waves and electron Bernstein waves.  Wave intervals range in duration from a few seconds to hours.  Both the amplitudes and {\color{black} number of detections} of these near-$f_{ce}$ waves increase significantly with decreasing distance to the Sun, suggesting that they play an important role in the evolution of electron populations in the near-Sun solar wind.  Correlations are found between the {\color{black} detection} of these waves and properties of solar wind electron populations, including electron core drift, implying that these waves play a role in regulating the heat flux carried by solar wind electrons.  {\color{black} Observation} of these near-$f_{ce}$ waves is found to be strongly correlated with near-radial solar wind magnetic field configurations with low levels of magnetic turbulence.  A scenario for the growth of these waves is presented which implies that regions of low-turbulence near-radial magnetic field are a prominent feature of solar wind structure near the Sun.

% 250 word limit for the abstract 

\end{abstract}

%%%%%%%%%%%%%%%%%%
%  Keywords
%%%%%%%%%%%%%%%%%%

\keywords{Solar wind -- wave-particle interactions -- electrons populations -- inner heliosphere}


%% ------------------------------------------------------------------------ %%
%
%  TEXT
%
%% ------------------------------------------------------------------------ %%

%%%%%%%%%%%%%%%%%%
%\section{Introduction}
%%%%%%%%%%%%%%%%%%

\section{Introduction} 
\label{sec:intro}

Sunward of $\sim 50$ solar radii ($R_S$), Parker Solar Probe detects intervals of strong wave power near the electron cyclotron frequency ($f_{ce}$).  These intervals range in duration from a few seconds to several hours.  The observed waves are electrostatic up to the sensitivity of the FIELDS instrumentation (no measurable signature in the search coil magnetometer).  Most of the wave power is concentrated at and below the electron cyclotron frequency ($f_{ce}$), and these waves often show strong harmonics.  At least two distinct concurrent wave modes are preliminarily identified as electrostatic whistler-mode waves and electron Bernstein waves.   

The amplitude and {\color{black} number of detections} of these near-$f_{ce}$ waves increases significantly as Parker Solar Probe approaches the Sun, suggesting that they play a role in the evolution of electron populations in the near-Sun plasma environment. Correlations are observed between the {\color{black} detection} of these waves and properties of solar wind electrons, which are composed of a core, halo, and strahl (the population of electrons escaping the solar corona) \citep{Montgomery1968, Feldman1975, Pilipp1987, Maksimovic2005}.  These correlations suggest that these waves play a role in regulating the solar heat flux carried by electrons.  Finally, the {\color{black} detection} of these waves is found to be strongly correlated with the presence of low-turbulence radial solar wind magnetic fields.

%%%%%%%%%%%%%%%
%% Prior Observations in the solar wind
%%%%%%%%%%%%%%%

Previous observations of plasma waves in the solar wind near $f_{ce}$ focused on whistler-mode waves.  These waves were reported in three primary contexts: (i) associated with solar wind plasma boundaries such as shocks (e.g. \citep{Wilson2009, RamirezVelez2012} and references therein) and stream interaction regions \citep{Beinroth1981, Lin1998, Lengyel1996, Breneman2010}, (ii) associated with the turbulent cascade of magnetic field fluctuations (e.g. \citep{Lengyel1996, Bruno2013, Narita2016} and references therein), and (iii) present in the free solar wind \citep{Zhang1998, Lacombe2014, Stansby2016, Tong2019}.  

Waves in context types i and iii are narrowband (clear sinusoidal waveforms in time-domain data), whereas the waves in context type ii are a superposition of waves, without clear sinusoidal waveforms. We do not elaborate further on the whistler-mode waves in context types i or ii, because the near-$f_{ce}$ waves observed by Solar Probe are not limited to association with shocks or stream interaction regions, and they are narrowband rather than broadband.  

Whistler-mode waves in the free solar wind are often thought to be generated by electron temperature anisotropy and/or heat flux instabilities (those involving a beaming component) \citep{Gary2005, Shaaban2018}.  Many numerical studies have focused on this issue (e.g. \citep{Vocks2005, Saito2007, Seough2015} and references therein), and data analyses of the radial evolution of solar wind electron distribution functions are frequently interpreted in terms of whistler-mode waves driving electron scattering \citep{Walsh2013, Graham2017, Bercic2019}.  Further, observed correlations between whistler-mode wave {\color{black} detection} and properties of electron distribution functions have been reported \citep{Kajdic2016, Stansby2016, Tong2019}. 

%%%%%%%%%%%%%%%
%% PSP observations quite different from all of these 
%%%%%%%%%%%%%%%

The near-$f_{ce}$ wave observations from Parker Solar Probe presented here are distinct from the vast majority of those reported in these prior studies in several regards: (i) the wave frequencies are considerably higher, centered on 0.7 $f_{ce}$ or $f_{ce}$ in Parker Solar Probe data, compared to $0.1 < f / f_{ce} < 0.3$ reported in prior studies \citep{Lengyel1996, Lin1998, Moullard2001, Lacombe2014, Kajdic2016, Stansby2016, Tong2019}, (ii) the waves are electrostatic up to the sensitivity of the FIELDS data, whereas most prior studies identified whistler-mode waves using exclusively magnetic field data \citep{Beinroth1981, Lacombe2014, Kajdic2016, Tong2019}, (iii) the waves observed by Parker Solar Probe are both narrow band and frequently observed, {\color{black} observed} up to 30\% of the time when magnetic field conditions favorable to wave growth exist (prior studies of non-turbulence whistler-mode waves show much lower detection rates (e.g. \citep{Lacombe2014, Tong2019})), and (iv) the near-$f_{ce}$ waves observed by Parker Solar Probe often include electron Bernstein modes, which have previously been reported in the solar wind only near shocks \citep{Wilson2010} or in conjunction with the AMPTE Li ion release \citep{Baumgaertel1989}.   

In addition to the waves described here, Parker Solar Probe does observe electromagnetic whistler-mode waves near 0.1 $f_{ce}$ (simultaneously observed signatures in both electric and magnetic field data with right-handed near-circular polarization). However, these lower frequency waves are not nearly as prevalent as the waves near $f_{ce}$ do not show the strong correlations with respect to ambient magnetic field properties described here.  

A full understanding of the origins of the near-$f_{ce}$ waves in the near-Sun solar wind, their potential importance to the evolution of solar wind electrons, and the information they carry about the large-scale structure of the solar wind will require considerable study.  Here we begin that process by reporting the existence of these waves, their observed properties, and correlations with solar wind conditions.

%%%%%%%%%%%%%%%%%%
%\section{Data Set and Processing}
%%%%%%%%%%%%%%%%%%

\section{Data Set and Processing}
\label{sec:methods}

Parker Solar Probe is a NASA mission designed to explore the near-Sun plasma environment \citep{Fox2016}.  Its prime mission consists of 24 orbits of the Sun, with progressively decreasing perihelion distances. The first two orbits, data from which are reported here, have a perihelion distance of 35.68 solar radii ($R_S$).  The solar 'encounter' where all Parker Solar Probe instruments operate at their nominal cadences begins at $\sim55$ $R_S$, providing radial coverage over $\sim20$ $R_S$ for the first two orbits.  

The FIELDS instrument \citep{Bale2016} measures in-situ electric fields from DC to $\sim20$MHz and in-situ magnetic fields from DC to $\sim1$MHz.  Electric fields are measured using five sensors. Four of these are 2m whip antennas located in the plane of the spacecraft heat shield and extending outward from the heat shield. Opposing antennas are 180$^{\circ}$ apart and the two antenna pairs are 5$^{\circ}$ from orthogonal with each other.  The differential voltage measurements made by these sensor pairs are designated $V_{12}$ ( = $V_1$ - $V_2$) and $V_{34}$ ( = $V_3$ - $V_4$).  The fifth sensor is a $\sim21$ cm antenna located on the magnetometer boom, in the umbra of the heat shield, 3.08 m from the spacecraft bus.  Magnetic fields are measured by a fluxgate magnetometer (DC to $\sim146.5$ S/s on three orthogonal axes) and a search coil magnetometer ($\sim10 Hz$ to $\sim20$ kHz on three orthogonal axes, $\sim10 kHz$ to $\sim1$ MHz on one axis).  All magnetic field sensors are located along the 3.5 m magnetometer boom.  

This study utilizes power spectra calculated on-board by the FIELDS Digital Fields Board (DFB) \citep{Malaspina2016_DFB}.  During the first two solar encounters, the reported AC power spectra are the mean of 16 individual power spectra calculated during the first 1/8 of each New York second (NYs = $2^{17}$ / 150,000 $\approx$ 0.874 s \citep{Bale2016}). AC power spectra are calculated for 4 channels.  For the first solar encounter, these channels were $V_{12}$ and the three low-frequency $SCM$ axes. For the second solar encounter, these channels were $V_{12}$, $V_{34}$, $V_5$, and the single-axis high frequency SCM winding.  For the first two encounters, the power spectral data were configured have 56 pseudo-logarithmically spaced frequency bins.  

This study utilizes data from the SWEAP instrument suite \citep{Kasper2016}, including proton moments from the sunward-facing SPC Faraday cup and electron distribution function data from the SPANe electron analyzers on the ram and anti-ram faces of the Parker Solar Probe spacecraft.

%%%%%%%%%%%%%%%%%%
%\section{Analysis}
%%%%%%%%%%%%%%%%%%

\section{Analysis}
 \label{sec:analysis_1}

\begin{figure}[ht!]
 \center
  \includegraphics[width=420px]{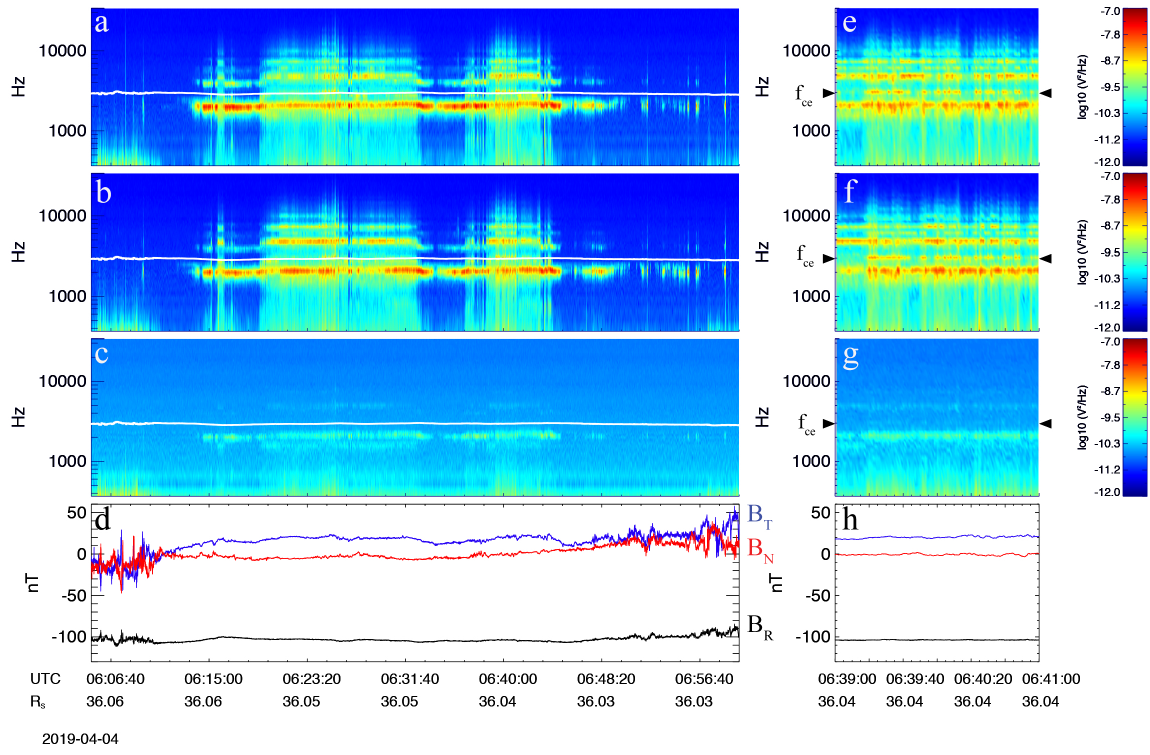}
 \caption{ Example of near-$f_{ce}$ waves in the near-Sun solar wind.  (a,b,c) Spectrograms of $V_{12}$ and $V_{34}$ differential voltage measurements and the $V_5$ single-ended voltage measurement, respectively.  Thick white lines indicate the local electron cyclotron frequency. (d) Ambient magnetic field vector in RTN coordinates.  {\color{black} (e,f,g,h) Spectrograms and magnetic field from 06:39 to 06:41 UTC, in the same format as (a,b,c,d).  Black arrows indicate $f_{ce}$ during these two seconds.} }
 \label{Fig_01}
 \end{figure}

Figure \ref{Fig_01} shows an example interval of near-$f_{ce}$ waves recorded on April 4, 2019.  The interval lasts $\sim30$ min.  Figures \ref{Fig_01}a, \ref{Fig_01}b, and \ref{Fig_01}c show AC power spectra for $V_{12}$, $ V_{34}$, and $V_5$.  The local value of $f_{ce}$ is indicated by a white line in each Figure.  Figure \ref{Fig_01}d shows the three components of the DC-coupled solar wind magnetic field in RTN coordinates.  These data were recorded near the second perihelion ($\sim36$ $R_S$).  The strongest wave power is generally observed at $\sim0.7$ $f_{ce}$ and significant harmonics are observed with amplitudes well above known instrumental response harmonics \citep{Malaspina2016_DFB}. During sub-intervals (e.g. near 06:39 or 06:24:00 UTC), the strongest wave power occurs near $\sim1.0$ $f_{ce}$, with significant harmonics. {\color{black} Because this can be difficult to discern in Figures \ref{Fig_01}a - \ref{Fig_01}d given the line indicating $f_ce$, Figures \ref{Fig_01}e - \ref{Fig_01}h show spectrograms and magnetic field data in the same format as Figures \ref{Fig_01}a - \ref{Fig_01}d, but focused on a short time interval from 06:39 to 06:41 UTC. Black arrows indicate $f_{ce}$ during this interval. Wave power at $f_{ce}$ and its harmonic is present. }  

The ambient solar wind magnetic field is disturbed prior to and after the wave interval, but is significantly less disturbed during the wave interval.  The solar wind magnetic field is also close to radial during the wave interval ($\sim10^{\circ}$ between the magnetic field vector and the direction radially outward from the Sun).  

%
%%%%%%%%
%% Figure 1
%%%%%%%%
%

To examine the degree to which near-$f_{ce}$ waves {\color{black} are observed} in association with near-radial solar wind magnetic field, an automated detection algorithm was applied to solar encounter 1 and 2 data to identify wave intervals.  First, each power spectra was converted to dB relative to background, where the background for each spectrum is defined as the median wave power at each frequency during the day when the observation was made.  Next, all peaks in each power spectra were identified using first and second derivatives (in the frequency dimension) of the power spectral data.  Power spectra where the largest peak was less than 8 dB above the noise were excluded from consideration.  Power spectra where the largest peak was less than 0.5 $f_{ce}$ or more than 1.1 $f_{ce}$ were also excluded from consideration.  Finally, a peak was identified as a potential near-$f_{ce}$ wave when the amplitude of the peak was at least 8 dB larger than the power of the first spectral point on either side of the peak where the derivative of the power spectra (in the frequency dimension) changed sign.  This algorithm was applied to the $V_{12}$ onboard AC spectral data only, as those data are available for both the first and second perihelion passes.  In all, 6.7 hours ($\sim$27,000 individual wave spectra) of data were found to contain near-$f_{ce}$ waves during solar encounter 1 and 9.12 hours ($\sim$37,000 individual wave spectra) during solar encounter 2.  

For each spectra identified as containing near-$f_{ce}$ waves, the angle between the solar wind magnetic field vector and the radial direction ($\theta_{Br}$) was calculated as the mean of the angle between the solar wind magnetic field vector and the radial direction over the NYs corresponding to each power spectral measurement. The sense of the radial field (sunward or anti-sunward) was not retained, such that the range of $\theta_{Br}$ is $0^{\circ} < \theta_{Br} < 90^{\circ}$. 
   
\begin{figure}[ht]
 \centering
 \includegraphics[width=250px]{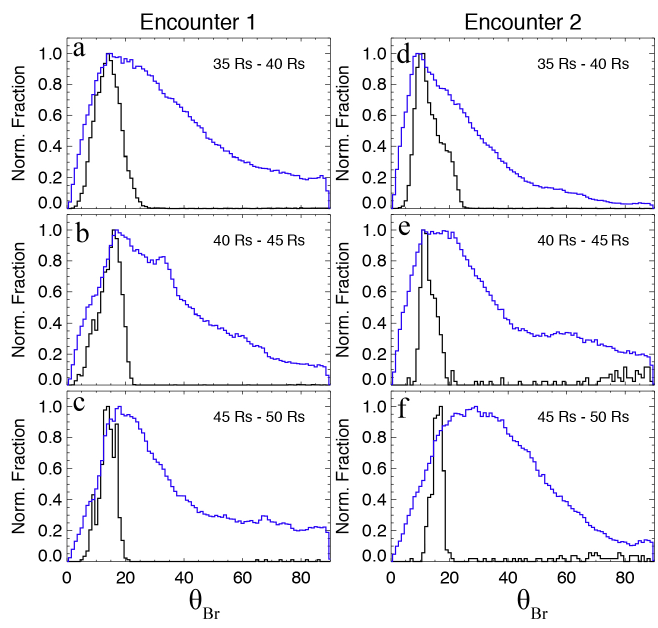}
 \caption{ Histograms of $\theta_{Br}$, the angle between the solar wind magnetic field vector and the radial direction. Data for when near-$f_{ce}$ waves were detected are indicated by black curves.  Data for all times are indicated by blue curves.  All curves are normalized to their maximum value.  (a,b,c) Histograms of $\theta_{Br}$ for solar Encounter 1, for the three indicated radial distance ranges.  (c,d,e) Same as (a,b,c), but for solar Encounter 2. }
 \label{Fig_02}
 \end{figure}

Figure \ref{Fig_02} compares the distributions of $\theta_{Br}$ for times when near-$f_{ce}$ waves were detected (black curves) and all solar wind (blue curves).  To enable comparisons, all distributions are normalized to their maximum value. Statistics for each solar encounter are shown as two columns (Figures \ref{Fig_02}a,\ref{Fig_02}b,\ref{Fig_02}c, and Figures \ref{Fig_02}d,\ref{Fig_02}e,\ref{Fig_02}f).  The data are divided into 5 $R_S$ segments.  In each case, the vast majority of near-$f_{ce}$ waves {\color{black} are detected} for $ \theta_{Br} < 25^{\circ}$.  Examining the few waves identified as near-$f_{ce}$ waves where $ \theta_{Br} > 25^{\circ}$ reveals that these are low-amplitude, isolated (to a single spectra) waves without harmonic signatures. From these data, we conclude that near-radial magnetic field is a necessary environmental condition for near-$f_{ce}$ wave growth.

The distributions with wave {\color{black} detections} (black curves) peak between $\sim$10$^{\circ}$ and $\sim$15$^{\circ}$, which approximately corresponds to the Parker spiral magnetic field deflection angle expected for undisturbed radially propagating solar wind by the time it reaches 35 and 55 solar radii, respectively (assuming 400 km/s solar wind).

%
%%%%%%%%
%% Figure 2
%%%%%%%%
%

\begin{figure}[ht]
 \centering
 \includegraphics[width=250px]{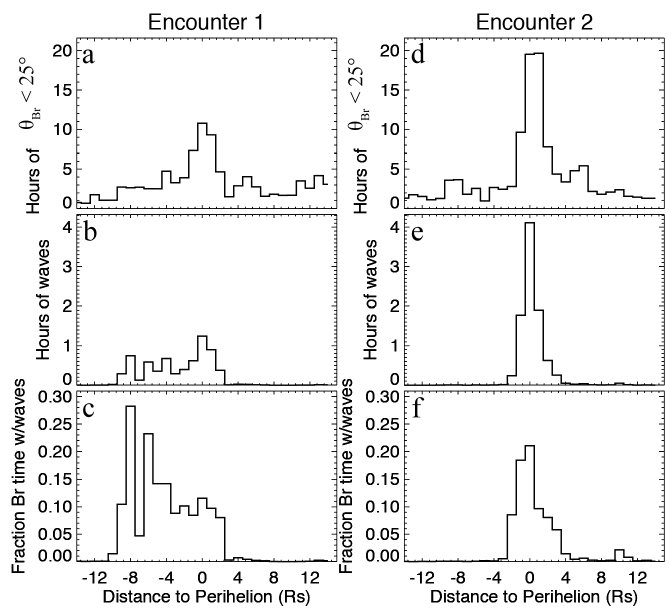}
 \caption{ Histograms of $\theta_{Br}$ and near-$f_{ce}$ wave {\color{black} detections} as a function of distance to perihelion. (a,d) Histogram of hours of data where $\theta_{Br} < 25^{\circ}$ for. (b,e) Histogram of hours of data where near-$f_{ce}$ waves are identified. (c,f) Histogram of the fraction of the solar wind magnetic field radial time when near-$f_{ce}$ waves were detected. (a,b,c) show data for solar encounter 1 and (d,e,f) for encounter 2. }
 \label{Fig_03}
 \end{figure}

Figure \ref{Fig_03} examines the variation in {\color{black} detection} of near-$f_{ce}$ waves with radial distance to the Sun.  Figures \ref{Fig_03}a and \ref{Fig_03}d show the number of hours where $\theta_{Br} < 25^{\circ}$ as a function of distance to perihelion (with 1 $R_S$ bins) for solar encounters 1 and 2, respectively.  Figures \ref{Fig_03}b and \ref{Fig_03}e show the number of hours when near-$f_{ce}$ waves were detected.  Figures \ref{Fig_03}c and \ref{Fig_03}f show the fraction of the solar wind magnetic field near-radial time when near-$f_{ce}$ waves were detected.  These data demonstrate that, when the magnetic field orientation is favorable for these waves to grow ($\theta_{Br} < 25^{\circ}$), they {\color{black} are observed} between 10\% and 30\% of the time.  

The radial profile of radial magnetic field observation is, to first order, symmetric with respect to the inbound and outbound motion of Parker Solar Probe for both solar encounters (Figures \ref{Fig_03}a, \ref{Fig_03}d).  On encounter 2, the {\color{black} observation} of near-$f_{ce}$ waves is likewise symmetric to first order (Figures \ref{Fig_03}e, \ref{Fig_03}f).  On encounter 1, the {\color{black} observation} of near-$f_{ce}$ waves was not radially symmetric (Figures \ref{Fig_03}b, \ref{Fig_03}c).  From these data, we conclude that near-radial magnetic field is a necessary, but not sufficient, environmental condition for near-$f_{ce}$ wave growth.  

%
%%%%%%%%
%% Figure 3
%%%%%%%%
%

\begin{figure}[ht]
 \centering
 \includegraphics[width=250px]{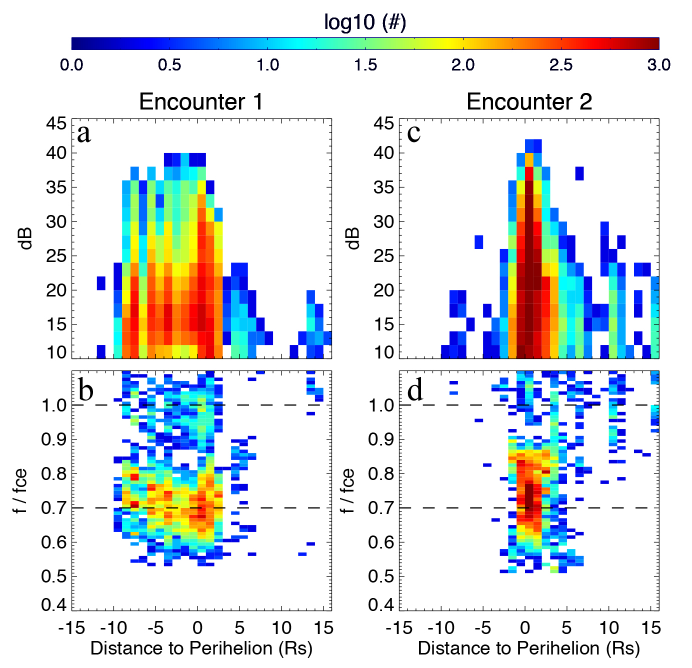}
 \caption{ Two dimensional histograms showing the number of near-$f_{ce}$ waves detected as a function of amplitude (in dB) or frequency (normalized to $f_{ce}$) and distance to perihelion. (a,b) show data for solar encounter 1.  (c,d) show data for solar encounter 2.  }
 \label{Fig_04}
 \end{figure}

Figure \ref{Fig_04} examines the variation in amplitude and frequency of near-$f_{ce}$ waves with radial distance to the Sun.  Figure \ref{Fig_04}a shows a two dimensional histogram of near-$f_{ce}$ wave {\color{black} detections} as a function of $V_{12}$ amplitude (in dB, of the largest spectral peak for each spectra) and distance, for solar encounter 1.  Figure \ref{Fig_04}c shows a similar plot for solar encounter 2.  For both encounters, the {\color{black} observation rate} of higher amplitude near-$f_{ce}$ waves increases toward perihelion, demonstrating that the waves become stronger and more frequent closer to the Sun.  The inbound / outbound symmetry in these figures follows that in Figure \ref{Fig_03}.  

Figure \ref{Fig_04}b shows a two dimensional histogram of near-$f_{ce}$ wave {\color{black} detections} as a function of wave frequency (frequency of the largest amplitude spectral peak for a given spectra) and distance, for solar encounter 1. Figure \ref{Fig_04}d shows a similar plot for encounter 2.  These data demonstrate that the presence of two wave modes inferred from case studies also appears in statistical analysis of the data.  Particularly in solar encounter 1, two distinct populations can be discerned, one with $0.6 f_{ce} < f < 0.8 f_{ce}$ and one with $0.9 f_{ce} < f < 1.1 f_{ce}$.  

%%%%%%%
%% Doppler Section 
%%%%%%%

%
%%%%%%%%
%% Figure 4
%%%%%%%%
%

\begin{figure}[ht]
 \centering
 \includegraphics[width=250px]{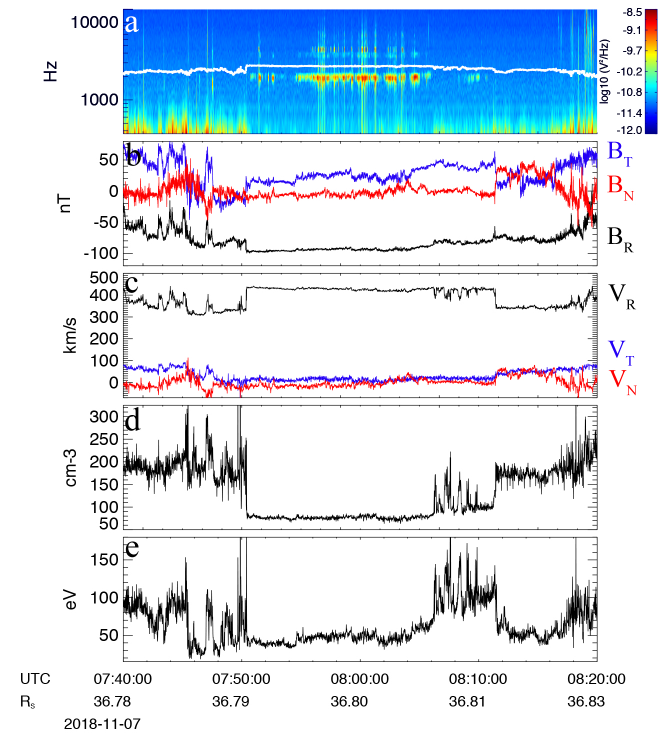}
 \caption{ Example of near-$f_{ce}$ waves in the near-Sun solar wind, with solar wind proton moments for context.  (a) Spectrogram of the $V_{12}$ differential voltage measurement. The thick white line indicates the local electron cyclotron frequency. (b) Ambient magnetic field vector in RTN coordinates. (c) Proton velocity in RTN coordinates. (d) proton density, (e) proton temperature. }
 \label{Fig_05}
 \end{figure}

Figure \ref{Fig_05} shows an interval of near-$f_{ce}$ waves along with proton moments as determined by SPC.  Figure \ref{Fig_05}a, \ref{Fig_05}b, \ref{Fig_05}c, \ref{Fig_05}d, and \ref{Fig_05}e show, respectively, the $V_{12}$  wave spectra, the DC-coupled magnetic field, solar wind velocity (spacecraft velocity removed), solar wind density, and proton temperature.  The period of wave activity is bracketed by distinct changes in solar wind conditions.  While these specific conditions (low density, faster solar wind speed) are not found to be broadly correlated with near-$f_{ce}$ wave {\color{black} detection}, these data do demonstrate that the solar wind associated with the near-$f_{ce}$ waves is distinctly different from the surrounding wind.  

With regard to \ref{Fig_05}e, it is important to note that SPC determines proton temperature in the sunward direction, such that the observed temperature variation may reflect a transition between measuring the temperature parallel to the background magnetic field (radial field) and measuring a combination of parallel and perpendicular temperatures (non-radial field). 

%
%%%%%%%%
%% Figure 5
%%%%%%%%
%

\begin{figure}[ht]
 \centering
 \includegraphics[width=250px]{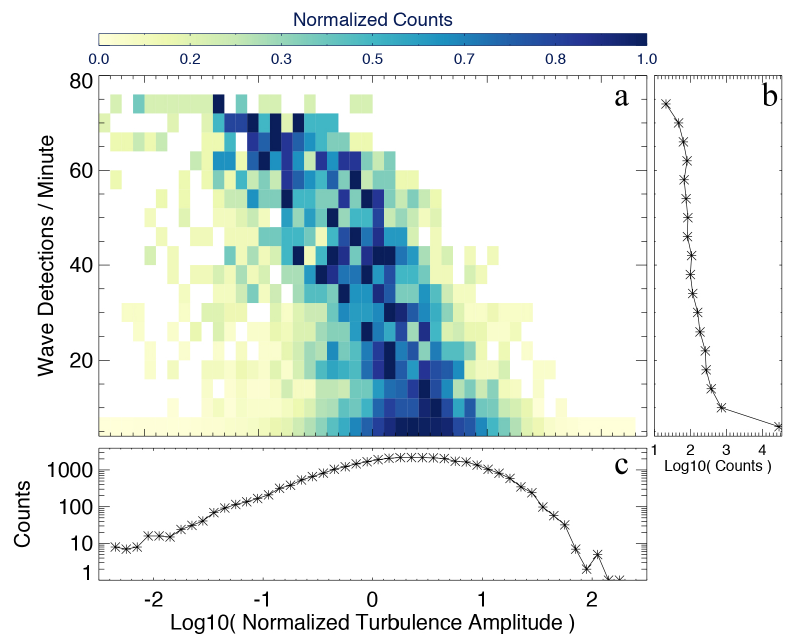}
 \caption{ (a) Two-dimensional histogram showing the number of 1-minute data samples, normalized to the maximum value in each horizontal row, as a function of near-$f_{ce}$ wave detections per minute and normalized magnetic field turbulence amplitude. (see text for explanations of the normalizations used).  (b) The number of 1-minute data samples in each row of the histogram in (a).  (c) The number of 1-minute data samples in each column of the histogram in (a). }
 \label{Fig_06}
 \end{figure}

To understand this difference in solar wind in a quantitative way, Figure \ref{Fig_06} explores the relationship between solar wind with low-amplitude magnetic fluctuations (`quiet' solar wind) and near-$f_{ce}$ waves.  Figure \ref{Fig_06}a shows a two dimensional histogram of wave {\color{black} detection} rate versus magnetic field turbulent amplitude.  Each horizontal row of this histogram has been normalized to the maximum value of counts in that row.  Figure \ref{Fig_06}b shows the number of counts (minutes of data with near-$f_{ce}$ wave detections) included in each row.  Figure \ref{Fig_06}c shows the number of counts (number of minutes) included in each column.  The maximum possible number of wave detections per minute exceeds 60 because the cadence of the spectral data used to detect the waves is $\sim0.87$ s (1 NYs).  

For this figure, the amplitude of the magnetic field turbulence is estimated as follows: a Fourier transform was calculated for each component of the DC-coupled magnetic field data within a minute-wide window.  The three component power spectral densities were summed.  The resulting spectra was flattened in frequency ($f$) space by dividing by $f^{-5/3}$. Finally, the average of the flattened power spectral density from 0.1 Hz to 1 Hz was calculated.  This is the turbulence amplitude value on the x axis in Figure \ref{Fig_06}.  A 1 Hz upper frequency cutoff is used to avoid the inclusion of ion cyclotron wave power \citep{Bale2019} and spacecraft-generated reaction wheel noise.   The data in Figure \ref{Fig_06} show a clear trend that near-$f_{ce}$ waves are preferentially {\color{black} observed} when the magnetic field turbulence is weak. 

%
%%%%%%%%
%% Figure 6
%%%%%%%%
%

%%%%%%%%%%%%%%%%%%
% Analysis of SPANe data 
%%%%%%%%%%%%%%%%%%

To gain insight into the origin of the instability that powers the near-$f_{ce}$ waves, we examine the {\color{black} detection} rate of near-$f_{ce}$ waves as a function of electron core sunward drift, in the proton frame, measured by the SPANe electron instrument on Parker Solar Probe.  

\begin{figure}[ht]
 \centering
 \includegraphics[width=250px]{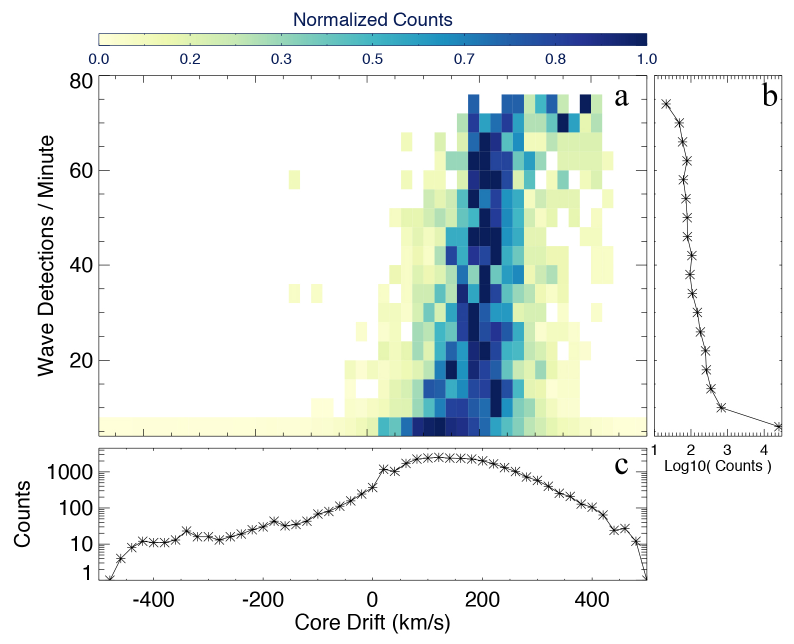}
 \caption{ (a) Two-dimensional histogram showing the number of 1-minute data samples, normalized to the maximum value in each horizontal row, as a function of near-$f_{ce}$ wave detections per minute and electron core drift velocity. (b) The number of 1-minute data samples in each row of the histogram in (a).  (c) The number of 1-minute data samples in each column of the histogram in (a). }
 \label{Fig_07}
 \end{figure}

Figure \ref{Fig_07} shows a two dimensional histogram in a similar format to Figure \ref{Fig_06}.  For Figure \ref{Fig_07}a, the horizontal axis shows the electron core drift velocity averaged over 1 minute intervals. Positive values indicate magnetic field-aligned drift, negative values indicate anti-field-aligned drifts. Because the background magnetic field was oriented near-radial (approximately sunward during identified near-$f_{ce}$ wave events) we interpret positive core drift velocities as sunward relative to the protons.  Data from solar encounters 1 and 2 are included here. Again, Figures \ref{Fig_07}b and \ref{Fig_07}c show the number of data samples included in each horizontal row and vertical column (respectively).  This analysis only includes electron distribution functions where the core could be fit (e.g. density detected by SPANe sufficiently high).  For details on the core fitting procedure and its limitations, see \citet{Halekas2019} (this issue).  

The core drift peaks near 100 km/s sunward for times when none or few near-$f_{ce}$ waves are observed (bottom row of Figure \ref{Fig_07}a).  As the number of waves detected per unit time increases, the core drift shifts toward 200 km/s, eventually reaching close to 300 km/s for the intervals with the most waves.  

Near-$f_{ce}$ waves {\color{black} are more frequently observed} in regions where the electron core drift is more strongly sunward (in the frame of the solar wind protons).  The sunward core drift velocity varies to balance the current associated with suprathermal electrons (strahl, halo) moving away from the Sun.  Because the halo fractional density drops significantly during the solar encounters \citep{Halekas2019} (this issue) and because the electron core and suprathermal currents nearly balance one another \citep{Halekas2019} (this issue), increases in the sunward core drift velocity indicate either: (i) regions where the strahl to core density increases, (ii) regions where the strahl velocity increases, or (iii) some combination of these.  Core drift is used as a proxy for strahl measurements here because the near-$f_{ce}$ waves {\color{black} are observed} most often when the background magnetic field is near radial and the strahl distribution is often partially or largely blocked from the SPANe sensors by the Parker Solar Probe heat shield.

%
%%%%%%%%
%% Figure 7
%%%%%%%%
%

%%%%%%%%%%%%%%%%%%
%\subsection{Discussion} and comparison with electron strahl widths  
%%%%%%%%%%%%%%%%%%

\section{Discussion}
\label{sec:res}

Near-$f_{ce}$ waves {\color{black} are observed} in regions where the ambient magnetic field is close to radial (Figure \ref{Fig_02}), magnetic turbulence is exceptionally weak (Figure \ref{Fig_06}), and the electron core sunward drift is enhanced (implying a larger suprathermal flux outward).  Further, \citet{Bale2019} reported that the near-Sun solar wind consists of `quiet' radial-field intervals with weak turbulent fluctuations, punctuated by intervals of Alfv\'{e}nic magnetic field reversals and strong magnetic field turbulence.  Given these observations, and the understanding of whistler-mode wave growth determined from theory, simulation, and space measurements prior to Parker Solar Probe (Section \ref{sec:intro}), we postulate the following scenario for the origin of the near-$f_{ce}$ waves.  

Flux tubes where magnetic field turbulence is low contain a larger outward flux of strahl electrons.  Those strahl electrons cause the sunward electron core drift (in the proton frame) to increase.  The combination of larger strahl flux and more sunward electron core drift set up electron distribution functions unstable to near-$f_{ce}$ wave growth.  Details of the specific instability and wave growth process will be explored in future work. 

The concept that the near-Sun solar wind is divided into 'quiet' magnetic flux tubes (where near-$f_{ce}$ waves are preferentially {\color{black} observed}) and 'strong turbulence' flux tubes where wave growth is suppressed is further supported by Figure \ref{Fig_01} and Figure \ref{Fig_05}, where the bulk of near-$f_{ce}$ wave power is observed in the center of each quiet, near-radial magnetic field region, rather than near the edges.  Waves near the edges would suggest growth due to instabilities associated with mixing plasma populations (e.g. \citep{Malaspina2015, Holmes2018}).  Waves near the magnetic structure center suggest that a property of the plasma within the flux tube is driving the instability.  

However, this picture is incomplete.  Why should flux tubes with 'quiet' solar wind (lower magnetic turbulence, hewing closer to the Parker spiral direction) show larger strahl electron flux?   Perhaps this indicates multiple coronal source region properties.  Perhaps it indicates different strahl radial evolution (efficiency of focusing and/or scattering) on 'quiet' versus 'strongly turbulent' magnetic flux tubes.  Future work will explore these possibilities.  

{\color{black} The preliminary identification of wave modes presented here (whistler / electron Bernstein) is based on wave frequency alone at this time.  Future detailed study of polarization and other wave parameters may require that these initial identifications be reconsidered.  
}

{\color{black} The correlation between near-$f_{ce}$ waves and electron core drift in Figure \ref{Fig_07} suggests that the waves are responding to changes in the local electron distribution function and supports the idea that the waves are generated close to where they are observed.  However, given the low-turbulence, near-radial magnetic field configurations where the waves are observed, it is possible that the waves are generated closer to the Sun and propagate to the observing spacecraft.  The efficiency of such a process could be enhanced if abrupt plasma density transitions at the boundaries of the low-turbulence, near-radial magnetic field regions (as in Figure \ref{Fig_05}) are common or persistent with radial distance from the Sun.  
}

Finally, the observations presented here demonstrate that near-$f_{ce}$ wave amplitude and occurrence in the data continue to increase toward the 35 $R_S$ perihelion (Figure \ref{Fig_04}), suggesting that regions of quiet radial-field solar wind will become ever more common as Parker Solar Probe reduces its perihelion distance on future orbits.  

%%%%%%%%%%%%%%%%%%
%\subsection{Conclusions}
%%%%%%%%%%%%%%%%%%

\section{Conclusions}
\label{sec:disc}

We presented observations of plasma waves near the electron cyclotron frequency sunward of 50 $R_{S}$.  These waves have frequencies centered near 0.7 $f_{ce}$ and $f_{ce}$, with strong harmonics.  They are electrostatic up to the sensitivity of FIELDS.  Their occurrence in the data and amplitude increase with decreasing distance to the Sun.  They {\color{black} are observed} during solar wind with near-radial magnetic field and weak magnetic field turbulence.  A scenario for wave growth was postulated based on enhancements in strahl electron flux and corresponding electron core drift enhancements. Supporting evidence for this scenario based on electron observations was presented. 

The study of near-$f_{ce}$ waves in the near-Sun solar wind has only just begun, and already it promises to provide insight into the regulation of electron heat flux (through improved understanding of electron population evolution and its connection with wave growth), the large-scale structure of the solar wind (by implying flux tubes of weakly turbulent magnetic field stretching back toward the Sun), and the nature of kinetic wave-particle interactions in the near-Sun solar wind.

%%%%%%%%%%%%%%%%%
%  ACKNOWLEDGMENTS
%%%%%%%%%%%%%%%%%

\acknowledgments

The authors thank the Parker Solar Probe team, especially the FIELDS and SWEAP teams for their support.  The FIELDS experiment on the Parker Solar Probe spacecraft was designed and developed under NASA contract NNN06AA01C. The authors wish to acknowledge helpful conversations with Dr. Ivan Vasko.    %All data used in this work are available on the FIELDS and SWEAP data archive: {\color{red} XX}.

%%%%%%%%%%%%%%%%%
%  Facilities 
%%%%%%%%%%%%%%%%%


\vspace{5mm}
\facilities{Parker Solar Probe (FIELDS, SWEAP)}


\software{SPEDAS \citep{Angelopoulos2019}  }

\end{document}